\documentclass[a4paper,10pt]{article}
\usepackage[utf8x]{inputenc}
\usepackage{tocloft}
\usepackage{parskip}
\usepackage{float}
\usepackage{upgreek}
\usepackage{amsmath}
\usepackage{amsthm}
\usepackage{amssymb}
\usepackage{multicol}
\usepackage{graphicx}

\title{A stabilised fiber-optic Mach-Zehnder interferometer for carrier-frequency rejection}
\author{Nathan Cooper$^1$, Jonathan Woods$^1$, James Bateman$^1$,\\Alexander Dunning$^1$ and Tim Freegarde$^1$ \\
\small{$^1$School of Physics and Astronomy, University of Southampton,}\\ \small{Highfield, Southampton SO17 1BJ, UK}}

\begin{document}

\maketitle

\begin{abstract}

We have demonstrated stabilisation of a fiber-optic Mach-Zehnder interferometer, with a centimetre-scale path difference, to the transmission minimum for the carrier wave of a frequency-modulated laser beam. A time-averaged extinction of 32~dB, limited by the bandwidth of the feedback, was maintained over several hours. The interferometer was used to remove the carrier wave from a 780~nm laser beam that had been phase-modulated at 2.7~GHz.

\end{abstract}

\begin{center}
 Copyright 2013 Optical Society of America
\end{center}

\hspace{1.5cm} \emph{OCIS codes:} 140.3298, 020.0020, 020.2930, 060.2310.

\section{Introduction}

Radio-frequency modulation allows the frequency precision of a stabilised laser to be coherently translated to adjacent wavelengths, and pairs of optical frequencies to be generated with low differential frequency noise. When achieved by electro-optic modulation, or direct modulation of the current through a diode laser, however, the different frequencies continue to propagate in the same beam mode. Spatial separation hence requires a frequency-dependent transmission or deflection. A commonly-used device for this is the Mach-Zehnder interferometer (MZI): with an appropriate path difference within the device, the sidebands emerge from one output port and the carrier from the other. For certain wavelengths, a similar effect can be achieved by employing the frequency-dependent Faraday effect in an atomic vapour \cite{reviewer1}.

Mach-Zehnder interferometers find applications as filters and frequency-dependent beam-splitters and combiners \cite{hbrick} for a variety of applications in optics \cite{fmziopt}, metrology \cite{fmzimet} and quantum information \cite{fmziqi}. Fiber-optic interferometers offer several advantages over free-space devices, such as improved output beam quality, easier integrability with other fiber-optic systems, lower sensitivity to vibrations and air currents, and a generally higher extinction ratio thanks to the ease and accuracy with which the beams may be recombined. Previous applications have included passively-stable microscopic devices for refractive index measurement \cite{microfibermzione}, interferometers that intentionally remain unstabilised for sensing purposes \cite{fmzimet,fmzimet2}, and devices that have been actively stabilised by dithering \cite{dither, rammach} or locking to the side of an interference fringe \cite{sidelock}. Polarization-dependent methods such as H\"ansch--Couillaud locking, which we have previously used to stabilise a free space interferometer \cite{machzen}, cannot easily be applied to fiber-optic devices because of the high temperature-dependence of the birefringence upon which polarization-maintaining fibers depend. Here, we demonstrate a similarly robust alternative stabilisation method that is suitable for fiber-based interferometers, using a convenient variation of frequency-modulation stabilisation \cite{pdh}.

\section{Interferometer stabilisation}

The stabilisation of an interferometer to maximize or minimize the transmission at a given wavelength presents the difficulty that a deviation to either side of the optimum configuration has the same effect upon the transmitted intensity; this parameter itself therefore gives no indication of the sign of the correction required. One solution is to perturb the interferometer in a slow, periodic `dither' and monitor the relative phase of variation of the transmitted intensity. A popular alternative is to accept a small offset and stabilise the transmitted intensity to a value just below its peak or above its minimum. In both cases, being near a turning point in the measured parameter, the interferometer must be significantly perturbed or displaced for a measurable signal to be obtained.

A more elegant technique, commonly used for the spectroscopic stabilisation of lasers and optical resonators, is to generate sidebands at frequencies either side of the wavelength required and measure the differential effect upon these components. In contrast to the large, slow perturbations of a dither, these components may be generated by the weak but rapid variations of frequency or phase modulation. The system remains in its stabilised configuration, and most of the optical power remains at the desired wavelength. Phase-sensitive detection of the beat note in the transmitted intensity then yields the error signal directly \cite{pdh}.

Here, partly for practical expedience and to avoid parasitic amplitude modulation, we use a variation upon conventional frequency modulation stabilisation and generate the sidebands by acousto-optic modulation of a fraction of the incident light, which we then direct backwards through the interferometer via the spare output port of the instrument. Where true frequency modulation uses beating of the sidebands with the carrier to provide a periodic alternation between the two sidebands, we achieve this directly by using a suitably aligned chopper wheel: although somewhat inelegant, this approach allows a large sideband spacing while requiring only low frequency detection circuitry. The derived error signal is used to control the temperature, and hence refractive index, of one arm of the interferometer via a thermoelectric cooler (TEC) upon which one of the fibers is mounted; it would alternatively be possible to stretch the fiber using a piezoelectric actuator, as in \cite{dither}.

\begin{figure}[H]
\begin{center}
\includegraphics[scale=0.70]{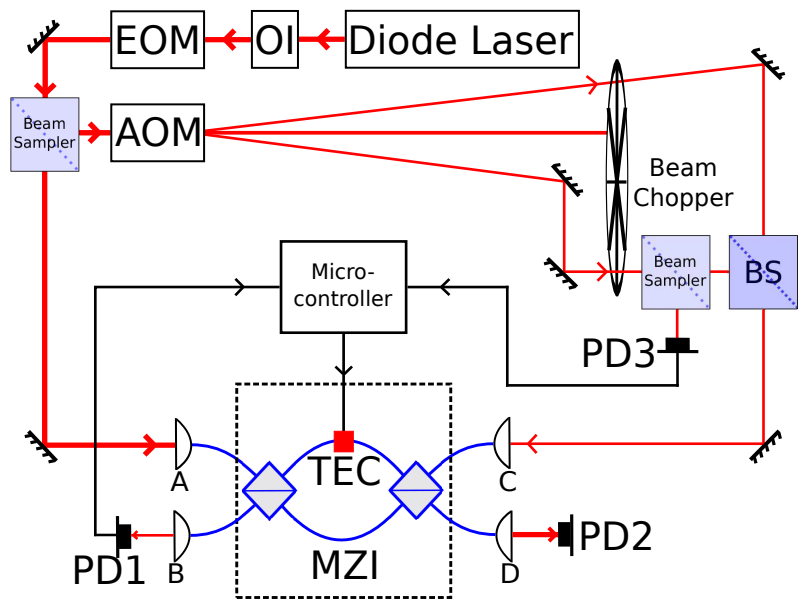}
\caption{(colour online) Experimental setup of the Mach-Zehnder Interferometer (MZI) and surrounding optics. BS: beam splitter, AOM: acousto-optical modulator, OI: optical isolator, TEC: thermoelectric cooler. Any of the light entering port A that emerges from port C is discarded, with unwanted feedback to the source laser being prevented by the optical isolator.}
\label{mzisetup}
\end{center}
\end{figure}

Our apparatus is shown schematically in figure \ref{mzisetup}. The Mach-Zehnder interferometer is a custom device, supplied by OzOptics \cite{ozopt}, with internal fiber arms 0.3~m long, one of which is attached to a thermoelectric cooler to regulate its temperature. An incident laser beam with frequency $\omega_0$, bearing sidebands at $\pm$~2.7 GHz, is directed into port A of the Mach-Zehnder interferometer; the sidebands are intended to emerge from port D, while the rejected carrier should be sent to port C. A fraction of the incident beam is split off from the main path and sent to the acousto-optic modulator, which produces two spatially distinct beams with frequencies $\omega_0 \pm \omega_{\text{AOM}}$, where $\omega_{\text{AOM}} \sim 2\pi \times$ 80~MHz. These are alternately fed backwards through the interferometer to the photodiode PD1. Using an indication of the chopper wheel phase provided by PD3, a microcontroller determines the error signal, which is proportional to the difference in transmitted intensity between the two AOM beams, and adjusts the current supplied to the thermoelectric cooler accordingly.

The error signal is given theoretically as follows. If the fiber lengths within the instrument are $l_0 \pm \Delta l/2$, and the fibers have a frequency-dependent refractive index $\eta(\omega)$ (including waveguide dispersion), then the accrued phase difference $\delta(\omega)$ between the two paths will be

\begin{equation}
 \delta(\omega) = \omega \eta(\omega) \Delta l / c
\end{equation}

and the difference in phase difference $\Delta\delta$ between optical frequencies $\omega_0$ and $\omega_0+\omega_m$ will hence be

\begin{equation}
 \Delta\delta(\omega_m) = \delta(\omega_0+\omega_m)-\delta(\omega_0) = (\Delta l/c) [(\omega_0+\omega_m) \eta(\omega_0+\omega_m) - \omega_0 \eta(\omega_0)],
\end{equation}

which may be expanded, assuming constant dispersion, to give

\begin{equation}
 \Delta\delta(\omega_m) \approx \frac{\Delta l}{c} \left[\omega_m \eta(\omega_0) + \omega_0 \frac{d\eta}{d\omega} \omega_m\right]
 = \frac{\Delta l \omega_m}{c} \left[\eta(\omega_0) + \omega_0 \frac{d\eta}{d\omega}\right]
 = \frac{\pi \omega_m}{\omega_s},
 \label{disp}
\end{equation}

where $\omega_s$ is the difference between frequencies that will be separated by the interferometer and is given by

\begin{equation}
 \omega_s = \frac{\pi c}{ \Delta l \left[\eta(\omega_0) + \omega_0 \frac{d\eta}{d\omega}\right]}.
 \label{omegas}
\end{equation}

The fraction of light transmitted between diagonally opposite ports of the interferometer will then be proportional to $\left|1+e^{i\delta(\omega)}\right|^2$, and the error signal will hence be proportional to

\begin{equation}
 \left|1+e^{i\delta(\omega_0+\omega_{\text{AOM}})}\right|^2 - \left|1+e^{i\delta(\omega_0-\omega_{\text{AOM}})}\right|^2,
\end{equation}

where $\delta(\omega\pm\omega_{\text{AOM}}) = \delta(\omega_0) \pm \Delta\delta(\omega_{\text{AOM}})$ as defined in equation (\ref{disp}) above.The form of the error signal $E_{s}$ is hence given by

\begin{equation}
 E_{s} \propto \left|1+e^{i(\delta(\omega_{0})+\pi \omega_{\text{AOM}}/\omega_{s})}\right|^{2} -  \left|1+e^{i(\delta(\omega_{0})-\pi \omega_{\text{AOM}}/\omega_{s})}\right|^{2}.
\end{equation}

\begin{figure}[H]
\begin{center}
\includegraphics[width=8cm, angle = 270]{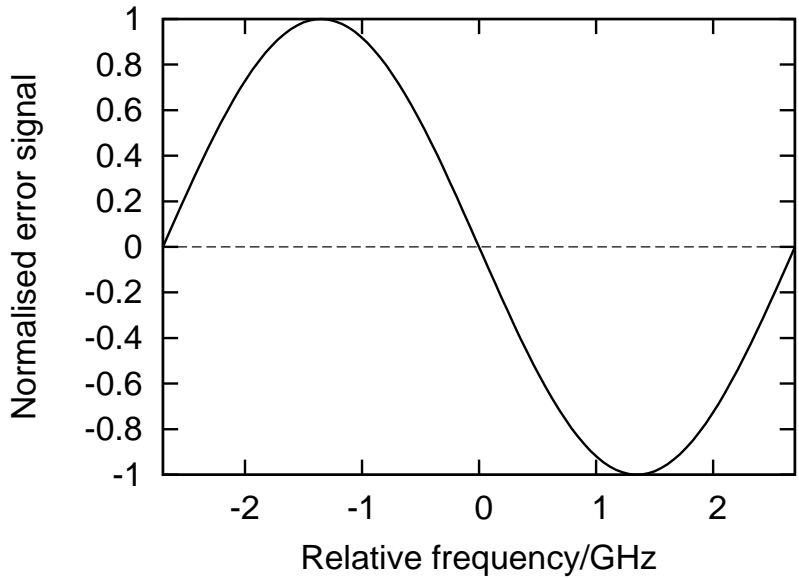}
\caption{Theoretical form of the error signal as a function of the difference in frequency between the (unshifted) input light and an optical frequency that would minimise $T_{AD}$, with $\omega_{\text{AOM}}/2\pi = 80$~MHz and $\omega_{s}/2\pi = 2.7$~GHz. Note that, as $\omega_{\text{AOM}} \ll \omega_{s}$, the error signal closely approximates a sinusoid.}
\label{theoplot}
\end{center}
\end{figure}

This is equal to zero and has finite gradient at both $\delta(\omega_0) = 0$ and $\delta(\omega_0) = \pi$. It is therefore suitable for stabilising the interferometer to maximize transmission of the carrier to either output port according to the sign of the feedback \cite{preextract}. A plot of this signal as a function of the frequency of the input beam is shown in figure \ref{theoplot}.

\section{Experiment and Results}

\subsection{Carrier removal from a phase-modulated spectrum}
\label{carrem}

\begin{figure}[H]
\begin{center}
\includegraphics[scale=0.5]{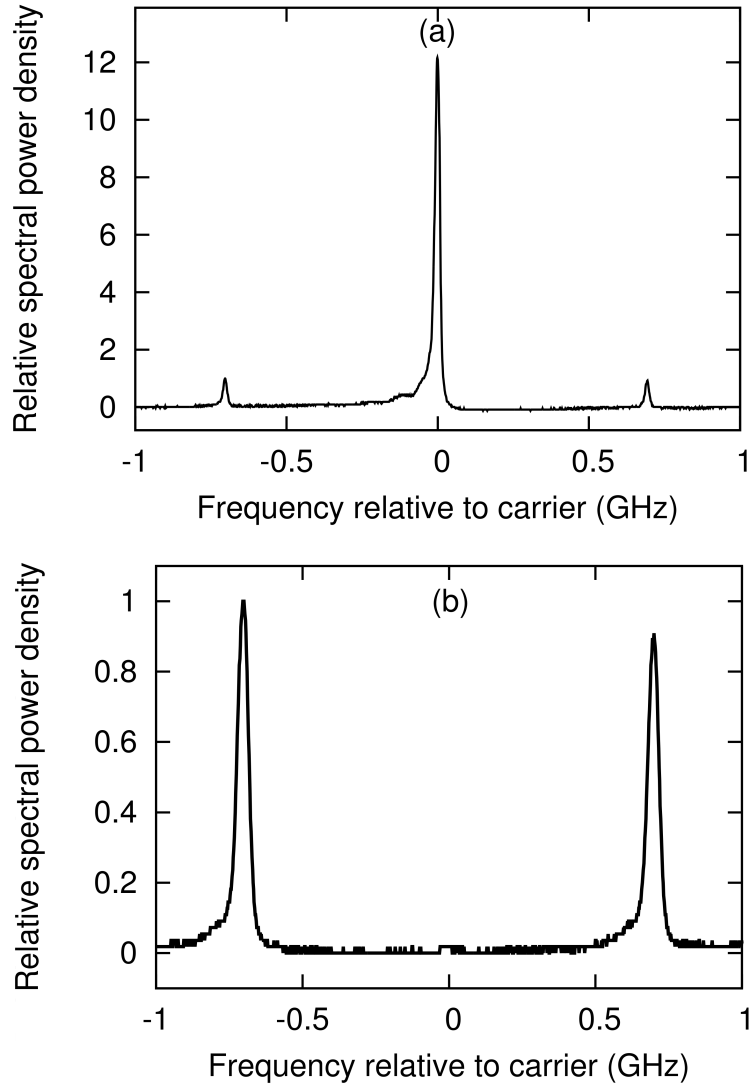}
\caption{Spectra of phase-modulated light before (a) and after (b) being passed through the interferometer while it was at a position of minimum carrier transmission to the relevant output. The modulation frequency is 2.7~GHz and our spectrum analyser is based on a cavity with a free spectral range of 2~GHz, hence the apparent appearance of the first order sidebands at $\pm 700$~MHz relative to the carrier. Interferometric suppression of the carrier is in excess of 30~dB.}
\label{sidebandplot}
\end{center}
\end{figure}

The separation of two closely spaced frequency components of a phase-modulated laser beam is a function that is often required in atomic physics experiments \cite{atomphys} and laser stabilisation schemes \cite{laserlock}. For our atomic physics experiments, for example, in which we stimulate Raman transitions between the ground hyperfine states of $^{85}$Rb, we modulate a 780~nm laser beam by passage through a 2.7~GHz electro-optic phase modulator (the remaining 300~MHz being provided by subsequent acousto-optic modulation), giving the spectrum shown in figure \ref{sidebandplot}(a), in which $\sim$15\% of the power is split between the two first-order sidebands. For use in our experiments the central carrier must be largely eliminated. Many previous solutions incur a significant power loss \cite{polfilt} or offer only a limited extinction ratio \cite{rammach}. 

The Mach-Zehnder interferometer allows the efficient separation of components that differ in frequency by an odd multiple of $\omega_s$, defined in equation (\ref{omegas}). With an optical path difference $\eta \Delta l$ specified for our device as 56.0$\pm$0.5~mm, $\omega_s$ = 2.7~GHz, and we may reject the carrier and send only the sidebands to the chosen output port of our interferometer, resulting in the spectrum shown in figure \ref{sidebandplot}(b), in which the carrier has been suppressed by over 30 dB through interferometric means. In addition to this, the carrier and sidebands are both attenuated by $\sim$4~dB by passage through the interferometer, owing to the poor beam quality of the incident light which limited the coupling efficiency into the interferometer. Although a path difference tolerance was specified for the practical device, no measurable difference was found between the positions of minimum carrier transmission and maximum transmission of the first-order sidebands.

\subsection{Feedback stabilisation of the inter-arm path difference}

The system in figure \ref{mzisetup} was constructed and the signals from each photodiode were sent to a simple digital control system based around an Arduino microcontroller \cite{PWM}. The interferometer itself was constructed to our specifications by OzOptics \cite{ozopt}, with the inter-arm path difference set such that $\omega_s = 2 \pi \times$ 2.7~GHz. Although our specifications allowed small inequalities in power division at the beamsplitters, we in practice observe no measurable effects. With the input at port A and a constant total laser power, the signal from photodiode PD2 can be used to derive the proportion $T_{AD}$ of the optical power emerging from the interferometer's outputs that leaves via port D. Equal powers must be coupled into the interferometer from each of the frequency-shifted beams entering port C; failure to do so results in narrowing of the locking region and a slight alteration of the lock point.

\begin{figure}[H]
\begin{center}
\includegraphics[scale=0.45]{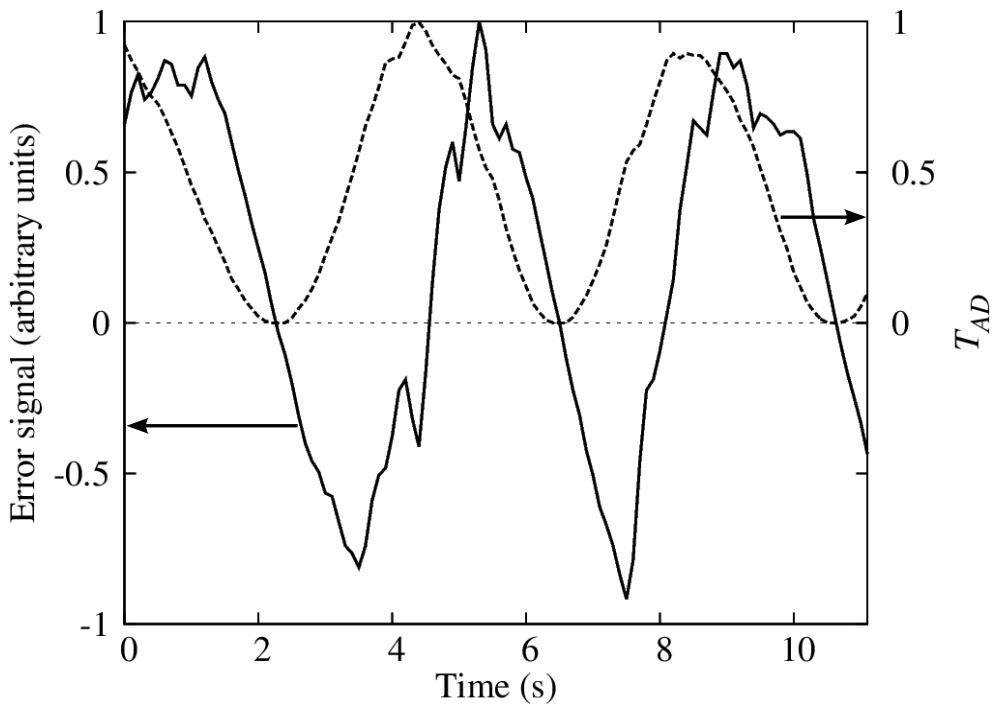}
\caption{Power transmission $T_{AD}$ from port A to port D, and error signal $E_s$, shown as functions of time during a rapid scan of the interferometer over several fringes. The noise on the error signal is primarily electrical in origin.}
\label{errvstransplot}
\end{center}
\end{figure}

Figure \ref{errvstransplot} shows the variation of the power transmission $T_{AD}$ and the error signal $E_s$ during a rapid scan of the inter-arm path difference in the interferometer, performed by making a sudden change to the current supplied to the TEC. The error signal not only exhibits a zero value corresponding with each minimum of $T_{AD}$, but also the desired antisymmetric form about these points.

\begin{figure}[H]
\begin{center}
\includegraphics[scale=0.6, angle = 270]{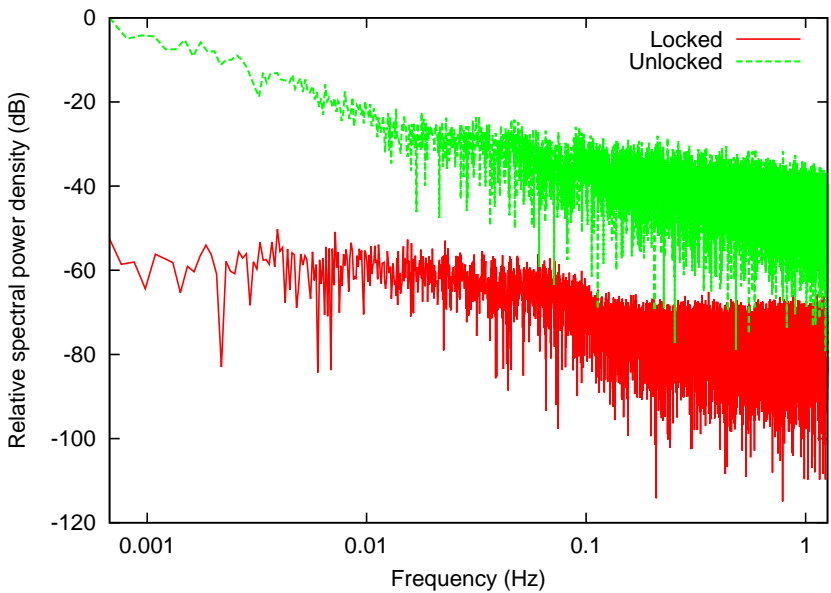}
\caption{Fast Fourier transforms of the measured values of $T_{AD}(t)$ over a period exceeding two hours during which the interferometer was either actively stabilised at a minimum of $T_{AD}$ (`locked') or allowed to drift freely (`unlocked'). Spectral power density values are given relative to the maximum value measured, which was found at 0.674~mHz with the interferometer unlocked.}
\label{FFT}
\end{center}
\end{figure}

The power transmission $T_{AD}$ was recorded for a period in excess of 2 hours, during which the interferometer was actively locked to a minimum of $T_{AD}$ (corresponding to $\delta_0 = \pi$). Fast Fourier transforms of $T_{AD}(t)$, over both this period and another period of equal length during which the interferometer was left unstabilised, are shown in figure \ref{FFT}. The time-average of $T_{AD}$ during this period was $6.3 \times 10^{-4}$ (corresponding to an interferometric extinction of 32.0~dB). We believe this to have been limited by the response rate of our feedback system, as measurably non-zero values of $T_{AD}$ were generally accompanied by a notable deviation of the error signal from zero. Extinction could therefore be improved by increasing the response rate of the feedback system, for example by replacing the TEC with a fiber-stretching piezoelectric transducer, as in \cite{dither}.  

\section{Conclusions}

We have demonstrated a stable locking method for a fiber-optic Mach-Zehnder interferometer with a macroscopic inter-arm path difference. Our device allowed the time-averaged output from a chosen port to be interferometrically suppressed by 32~dB for a period of over two hours. The extinction is believed to be limited by the response rate of our feedback system, which might be improved by switching from thermal to mechanical feedback. Our interferometer has been used to separate the carrier wave from the first-order sidebands of a 780~nm laser beam after electro-optic modulation at 2.7~GHz.

\section{Acknowledgements}

This work was supported by the UK EPSRC grants EP/E039839/1 and EP/E058949/1.


\begin{thebibliography}{99}
\bibitem{preextract}  In our experiment, the light used to generate the locking signal was taken from the laser beam before the electro-optic phase modulator, and hence contained only a single frequency component. If instead the sidebands are generated, for example, by modulation of the laser supply current, they will remain present in the monitor beam and could lead to additional features and false lock-points. However, provided that the sidebands are separated from the carrier by integer multiples of $\omega_s$, they will merely change the magnitude of the error signal and not alter its form, although in the case of strong modulation the sign of the error signal may be reversed.
\end{thebibliography}
\end{document}